\documentclass{article}
\usepackage{graphics}
\usepackage{graphicx}
\usepackage{epstopdf}
\begin{document}
\newcommand{\grad}{\mbox{\boldmath$\nabla$}}
\newcommand{\bdiv}{\mbox{\boldmath$\nabla\cdot$}}
\newcommand{\curl}{\mbox{\boldmath$\nabla\times$}}
\newcommand{\bcdot}{\mbox{\boldmath$\cdot$}}
\newcommand{\btimes}{\mbox{\boldmath$\times$}}
\newcommand{\btau}{\mbox{\boldmath$\tau$}}
\newcommand{\btheta}{\mbox{\boldmath$\theta$}}
\newcommand{\bphi}{\mbox{\boldmath$\phi$}}
\newcommand{\bmu}{\mbox{\boldmath$\mu$}}
\newcommand{\bepsilon}{\mbox{\boldmath$\epsilon$}}
\newcommand{\bcj}{\mbox{\boldmath$\cal J$}}
\newcommand{\bcf}{\mbox{\boldmath$\cal F$}}
\newcommand{\bbeta}{\mbox{\boldmath$\beta$}}
\newcommand{\lbar}{\lambda\hspace{-.09in}^-}
\newcommand{\bcp}{\mbox{\boldmath$\cal P$}}
\newcommand{\bcr}{\mbox{\boldmath$\cal R$}}
\newcommand{\bco}{\mbox{\boldmath$\omega$}}
\newcommand{\brho}{\mbox{\boldmath$\rho$}}
\title{Electric field of a point charge in truncated hyperbolic motion.}
\author{Jerrold Franklin\footnote{Internet address:
Jerry.F@TEMPLE.EDU}\\
Department of Physics\\
Temple University, Philadelphia, PA 19122}
\maketitle
\begin{abstract}
We find the electric field of a point charge 
in `truncated hyperbolic motion', in which the charge moves at a constant velocity followed by motion with
a constant acceleration in its instantaneous rest frame.  The same Lienard-Wiechert formula holds for the acceleration phase and the constant velocity phase of the charge's motion.  The only modification is that the formula giving the retarded time is different for the two motions, and the acceleration is zero for the constant velocity motion.
The electric field lines are continuous as the retarded time increases through the transition time between constant velocity and accelerated motion.  As the transition time approaches negative infinity the electric field develops a delta function contribution that has been introduced by others as necessary to preserve Gauss's law for the electric field.

\end{abstract}

\section{Introduction}
A point charge with a constant acceleration ${\bf a}_0=a_0{\bf{\hat z}}$ in its instantaneous rest system follows the relativistic trajectory\cite{gauss}
\begin{equation}
z_q(t)=\sqrt{t^2+1/a_0^2}.
\label{zt}
\end{equation}
  The equation relating $z_q$ and $t$ is a hyperbola, leading to the common designation of this motion as `hyperbolic motion'.
The trajectory in Eq.~(\ref{zt}) corresponds to a charge that comes to rest at 
$z_q=z_0=1/a_0 $ at time $t = 0 $ after traveling an infinite distance from the infinite past where its speed $\rightarrow $1.

The electric field produced by the accelerating charge has been calculated in Ref.~\cite{fg}.
The field lines show an unusual, non-physical behavior at $z=0$ (for observation time t=0)
where they abruptly stop.  This is because the retarded position of the charge is 
outside the past light cone for any $z<0$.
This problem is recognized  in \cite{fg}, and they attempt to correct it by introducing ``truncated hyperbolic motion", in which the accelerated part of the motion is preceded by  motion with a constant velocity.

However, their results in figures 3 and 4 of their paper show an unphysical discontinuous behavior for the electric field lines as the charge passes through the transition from constant velocity to accelerated motion.
In this paper, we modify the calculation of the electric field lines, resulting in continuous electric field lines that satisfy the appropriate physical constraints.

In section 2 of this paper, we summarize the calculation in \cite{fg} of the Lienard-Wiechert electric field for hyperbolic motion, and agree with their conclusion that the abrupt end of electric field lines at $z=0$ is unphysical.
In section 3, we derive the
electric field for the constant velocity phase of the motion and for the accelerated phase from the same 
standard Lienard-Wiechert formula, only using different retarded time formulas for the two different situations.
Consequently, our electric field lines are 
continuous as the retarded time increases through the transition time  between constant velocity and accelerated motion, and our figures 2 and 3 do not show the discontinuous behavior of the corresponding figures 3 and 4 in \cite{fg}. 
As the transition time approaches negative infinity the constant velocity part of the electric field develops a delta function contribution that has been proposed by others as necessary to preserve Gauss's law for the electric field.

\section{Lienard-Wiechert electric field for hyperbolic motion}

The Lienard-Wiechert electric field of a unit point charge is given by
\begin{equation}
{\bf E(r},t)=\left\{\frac{({\bf R}-R{\bf v})(1-v^2)
+{\bf R}\btimes[({\bf R}-R{\bf v)\btimes a}]}
 {(R-{\bf R\bcdot v})^3}\right\}_{\rm ret},\hspace{.3in}
\label{lw}
\end{equation}
where $\bf R$ is the vector from the unit point charge to the point of observation.
All variables (${\bf R,v,a}$)
on the right-hand side of Eq.~(\ref{lw}) are evaluated at the retarded time, $t_r=t-R$.

For the trajectory in Eq.~(\ref{zt}), the electric field 
is derived in Eqs.~(2)-(11) of \cite{fg}.  We summarize their derivation below.
The variables in Eq.~(2) are given (in cylindrical coordinates) by
\begin{eqnarray}
{\bf R}&=&\brho+[z-z_q(t_r)]{\bf\hat z}=
\brho+\left(z-\sqrt{z_0^2+t_r^2}\right){\bf{\hat z}}\\
{\bf v}&=&\frac{t_r{\bf{\hat z}}}{\sqrt{z_0^2+t_r^2}}\label{v}\\
{\bf a}&=&\frac{z_0^2{\bf{\hat z}}}{(z_0^2+t_r^2)^{\frac{3}{2}}}.
\end{eqnarray}

For simplicity, we evaluate the electric field at time $t=0$.
For $t=0$, the retarded time is negative and satisfies the relation
\begin{equation}
t_r^2=R^2=\rho^2+\left(z-\sqrt{z_0^2+t_r^2}\right)^2
=\rho^2+z^2-2z\sqrt{z_0^2+t_r^2}+z_0^2+t_r^2,
\label{tr}
\end{equation}
with the solution
\begin{equation}
t_r=\frac{-\sqrt{(\rho^2+z^2+z_0^2)^2-(2zz_0)^2}}{2z},\quad z\ge 0.
\label{tra}
\end{equation}
Putting Eqs.~(3)-(7)  into Eq.~(\ref{lw}) gives the electric field (after some algebra)
\begin{equation}
{\bf E}(\rho,z)=\frac{4z_0^2[(z^2-z_0^2-\rho^2){\bf{\hat z}}+2z\brho]}{[(\rho^2+z^2+z_0^2)^2-(2
zz_0)^2]^{\frac{3}{2}}}\Theta(z).
\label{Eh}
\end{equation}

The electric field must vanish for negative z because it can be seen from Eq.~(\ref{tr}) that there is no solution for $t_r$ for negative $z$.
Physically, this results from the fact that a charge with the trajectory given by Eq.~(\ref{zt}) is always outside the past light cone of any point with negative $z$ (at $t= 0$), 
and consequently cannot have any effect for negative z.	

The field lines for ${\bf E}(\rho,z)$ can be found from either Eq.~(\ref{lw})  or Eq.~(\ref{Eh}).  
They, of course, give the same answer because  Eq.~(\ref{Eh}) is derived from Eq.~(\ref{lw}). 
The electric field lines are
 plotted in Fig.~2 of \cite{fg}, and in Fig.~1 of this paper.   
\begin{figure}[h]
\begin{centering}
\includegraphics[height=4in]{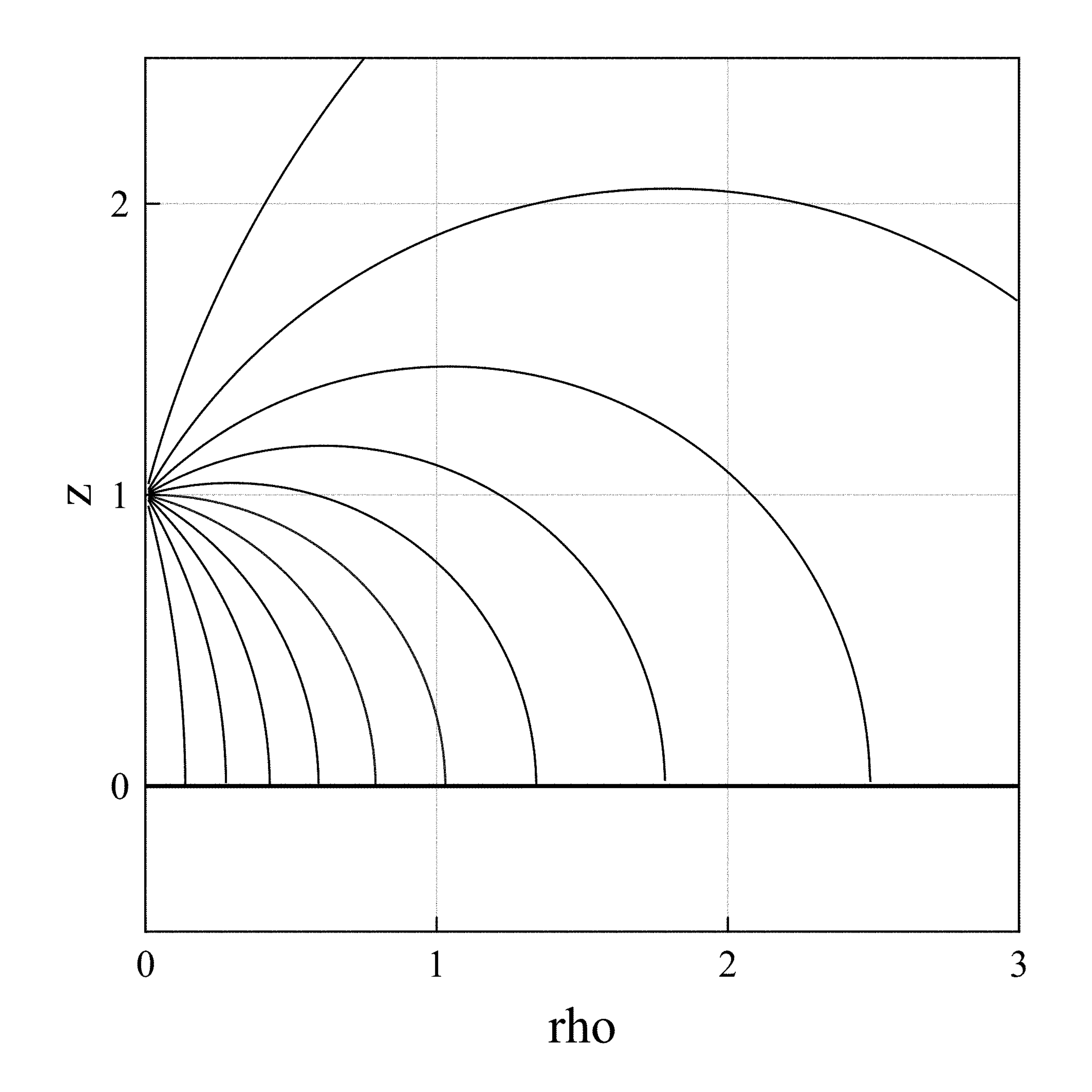}
\end{centering}
\caption{Electric field lines of a charged particle in hyperbolic motion at t=0.
The field lines are calculated from Eq.~(\ref{lw}), with the retarded time given by Eq.~(\ref{tra}).
The scale is set with $z_0=1$.}
\end{figure}
A remarkable feature of these curves is that the field lines stop abruptly at $z=0$
because there is no solution for the retarded time for negative $z$.
It is recognized in \cite{fg} that the field given by Eq.~(\ref{Eh}) and represented in our Fig.~1 cannot be a physical electric field,
because an electric field cannot just end in empty space without terminating on electric charge.  This would violate Maxwell's equation  $\nabla\cdot{\bf E}=4\pi\rho$, which leads to Gauss's law and the requirement that the normal component of E be continuous across any surface without a surface charge.
We attribute this failure of the Lienard-Wiechert electric field to the fact that the limit $v\rightarrow 1$ (even if in the distant past), required for complete hyperbolic motion, is unphysical in itself.  

\section{Electric field for truncated hyperbolic motion}

Following \cite{fg}, we modify the motion of the point charge so that it is originally moving at a constant velocity $\bf V$ until a time $T$, after which it moves
 with a constant acceleration ${\bf a}_0$ in its instantaneous rest system.  Reference \cite{fg} calls this motion ``truncated hyperbolic motion".
We continue to use Eq.~(\ref{lw}) for constant velocity by just setting $\bf a$ equal  $\bf 0$.  The only other change we have to make is to use the retarded time appropriate to the constant velocity motion.

For constant velocity motion up to the time $T$ at which the acceleration starts, the velocity is given by
\begin{equation}
{\bf V}=\frac{T{\bf\hat z}}{\sqrt{z_0^2+T^2}},\quad t< T,
\label{vv}
\end{equation}
which follows by just substituting $T$ into Eq.~(\ref{v}).  
The constant velocity trajectory of the charge is given by
\begin{equation}
z_q(t)=\frac{z_0^2+tT}{\sqrt{z_0^2+T^2}}=\frac{z_0^2+tT}{Z},\quad t<T,
\label{zqv}
\end{equation}
where we have introduced the notation $Z=\sqrt{z_0^2+T^2}$ for the position of the charge at the time the acceleration starts.
Then, the retarded time for the constant velocity motion of the charge satisfies the relation
\begin{equation}
t_r^2=R^2=\rho^2+\left(z-\frac{z_0^2+t_rT}{Z}\right)^2,\quad  t_r<T,
\end{equation}
with the solution
\begin{equation}
t_r=\frac{\left[(z_0^2-zZ)T-Z\sqrt{(zZ-z_0^2)^2+\rho^2 z_0^2}\right]}{z_0^2},\quad t_r<T.
\label{trv}
\end{equation}
Note that the restriction $z\ge 0$ no longer applies.

The electric field for the constant velocity part of the truncated hyperbolic motion is given by substituting Eqs.~(\ref{vv}), (\ref{zqv}), and (\ref{trv})
into Eq.~(\ref{lw}) with {\bf a} taken as $\bf 0$.
The result for ${\bf E}$ is given in [2] as\cite{dg}
\begin{equation}
{\bf E}=\frac{(1-V^2){ \bcr}}{[{\cal R}^2-V^2\rho^2]^{3/2}},\quad t\le T,
\label{Ev}
\end{equation}
where
\begin{equation}
{\bcr}={\brho}+{\bf z}-{\bf\hat z}z^2_0/Z
\label{bcr}
\end{equation}
is the vector to the point of observation
from the point 
\begin{equation}
z=z_0^2/Z
\end{equation}
on the z-axis which  the unit point charge would  have reached had it followed the constant velocity trajectory until the observation time, $t=0$.
Using Eq.~(\ref{bcr}), we can write Eq.~(\ref{Ev}) as
\begin{equation}
{\bf E}=\frac{z_0^2[{\brho}Z+{\bf\hat z}(zZ-z_0^2)]}{[\rho^2z_0^2+(zZ-z_0^2)^2]^{3/2}},\quad t\le T.
\label{Er}
\end{equation}

We have derived the electric field lines for all the figures of this paper by the following procedure:
We start close to the point $z_0$ at $15^o$ intervals from the z axis, where we calculate initial values $E_{\rho 0}$ and $E_{z 0}$.
Then we determine each next point in the field lines by
\begin{eqnarray}
\rho_{n+1}&=&\rho_{n}+E_{\rho n}\Delta/E_n\\
z_{n+1}&=&z_{n}+E_{z n}\Delta/E_n,
\end{eqnarray}
where $\Delta$ is a small increment (We chose $\Delta=0.01z_0$.).

The electric field values were calculated from Eq.~(\ref{lw}), using Eq.~(\ref{tra}) for $t_r\ge T$ and Eq.~(\ref{trv}) for $t_r\le T$.
We also used Eq.~(\ref{Eh}) For $t_r\ge T$ and Eq.~(\ref{Er}) for $t_r\le T$ as a check on our calculations.
The field lines of $\bf E$ for several values of T, the time at which the acceleration starts, are shown in figures 2, 3, and 4. These figures can be considered corrected versions of figures 3, 4, and 5 of \cite{fg}.  
\begin{figure}[h]
\begin{centering}
\includegraphics[height=4in]{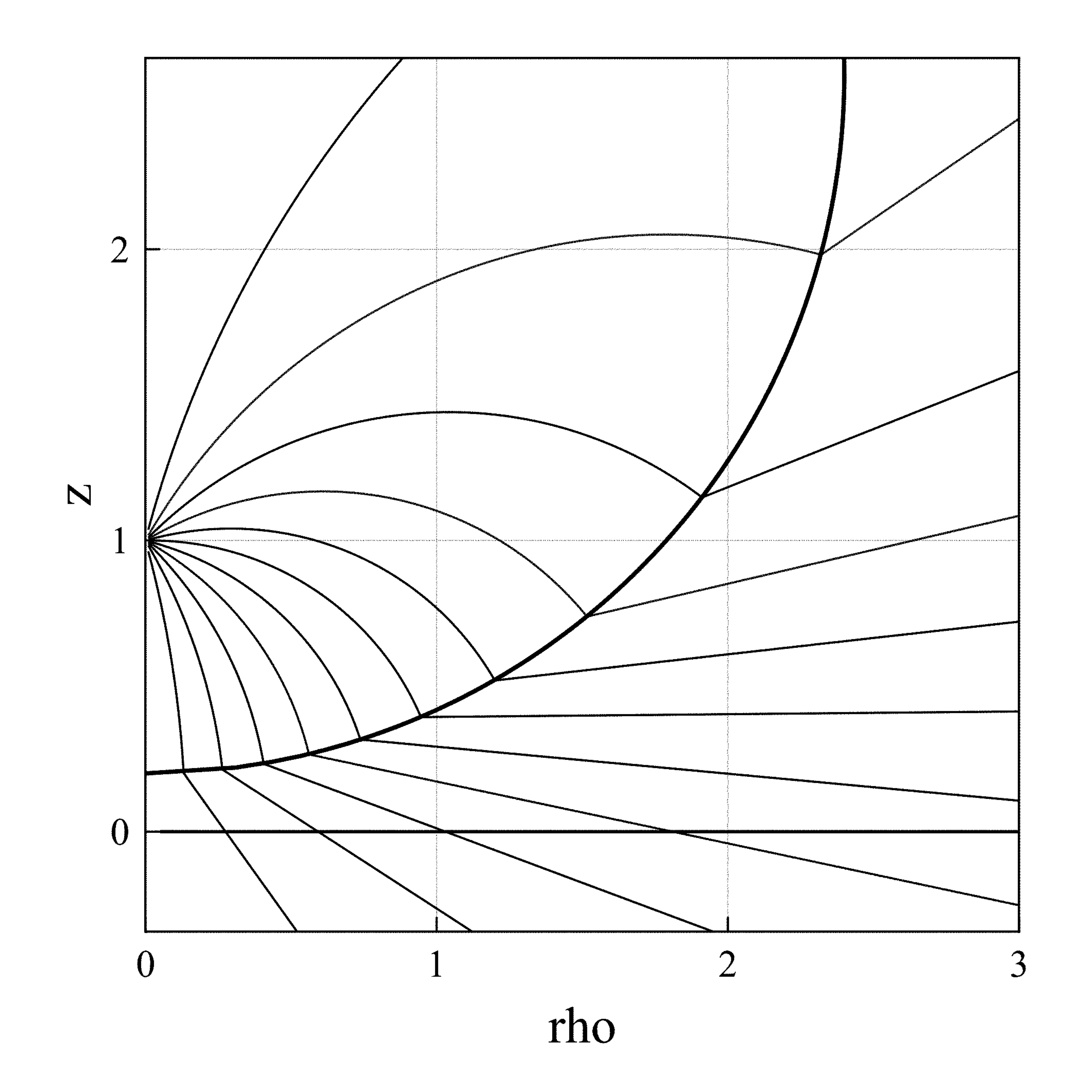}
\end{centering}
\caption{Electric field lines of a charged particle in truncated hyperbolic motion as given by Eq.~(\ref{lw})
with $T=-2.4$. The acceleration is zero for the lines outside the sphere  (shown as a circle on the figure).
The lines inside the sphere
 use Eq.~(\ref{tra}) for the retarded time, and those outside the sphere
use Eq.~(\ref{trv}).}
\end{figure}

In figure 2, and subsequent figures, the transition from constant velocity motion to accelerated motion takes place on a sphere defined by the equation
\begin{equation}
\rho^2+(z-Z)^2=T^2.
\label{circle}
\end{equation} 
The field lines outside the sphere are straight lines directed radially outward from the point $z_0^2/Z$ on the z-axis.
The acceleration electric field lines  are curved lines starting at the point $z_0$ on the z-axis, and ending at the transition sphere.
At the transition sphere they are in the $\bf\hat R$ direction, directed radially outward from Z, the retarded position of the charge.
This can be seen from the cross product of $\bf R\times E$, which vanishes just inside the sphere.
 
Figure 2 shows that the field lines are continuous through the transition from constant velocity motion to accelerated motion.  As the field lines pass through the spherical surface between the accelerated and the constant velocity motion, the normal component of $\bf E$ is continuous, and Gauss's law is satisfied at the surface.  This can be seen from Eq.~(\ref{lw}), where the discontinuity in the acceleration does not affect the radial component (${\bf\hat R}\bcdot{\bf E}$) of the retarded field. 
  This is in contrast to Fig.~3 of \cite{fg}, where the field lines inside and outside the sphere are disconnected, and the continuity of the normal component of $\bf E$ and Gauss's law are violated.     

An electric field does not usually have a kink in its field lines.  This is because Maxwell's equation 
$\curl{\bf E}=-\partial_t{\bf B}$ usually
requires that the transverse component of E be continuous as a field line crosses a surface.
This follows from Stokes' law which relates the discontinuity in the transverse component of $\bf E$ to the integral  $\int_{z_-}^{z_+}(-\partial_t{\bf B})dz$,
where $z_-$ and $z_+$ are just inside and just outside the transition sphere. 
The integral is zero if $\partial_t{\bf B}$ is finite, but in the present case $\partial_t{\bf B}$ is  infinite because the acceleration is discontinuous.
If we write the electric field $\bf E$ in terms of its  constant velocity and acceleration parts as 
$\bf E=E_v+ E_a$, then $\partial_t\bf B$ has a delta function part, given by
$\partial_t{\bf B={\hat R}\btimes E_a}\delta(z-z_-)$, so the integral equals ${\bf{\hat R}\btimes E_a}(z_-)$, which is just the discontinuity in the transverse component of $\bf E$.
\begin{figure}[h]
\begin{centering}
\includegraphics[height=4in]{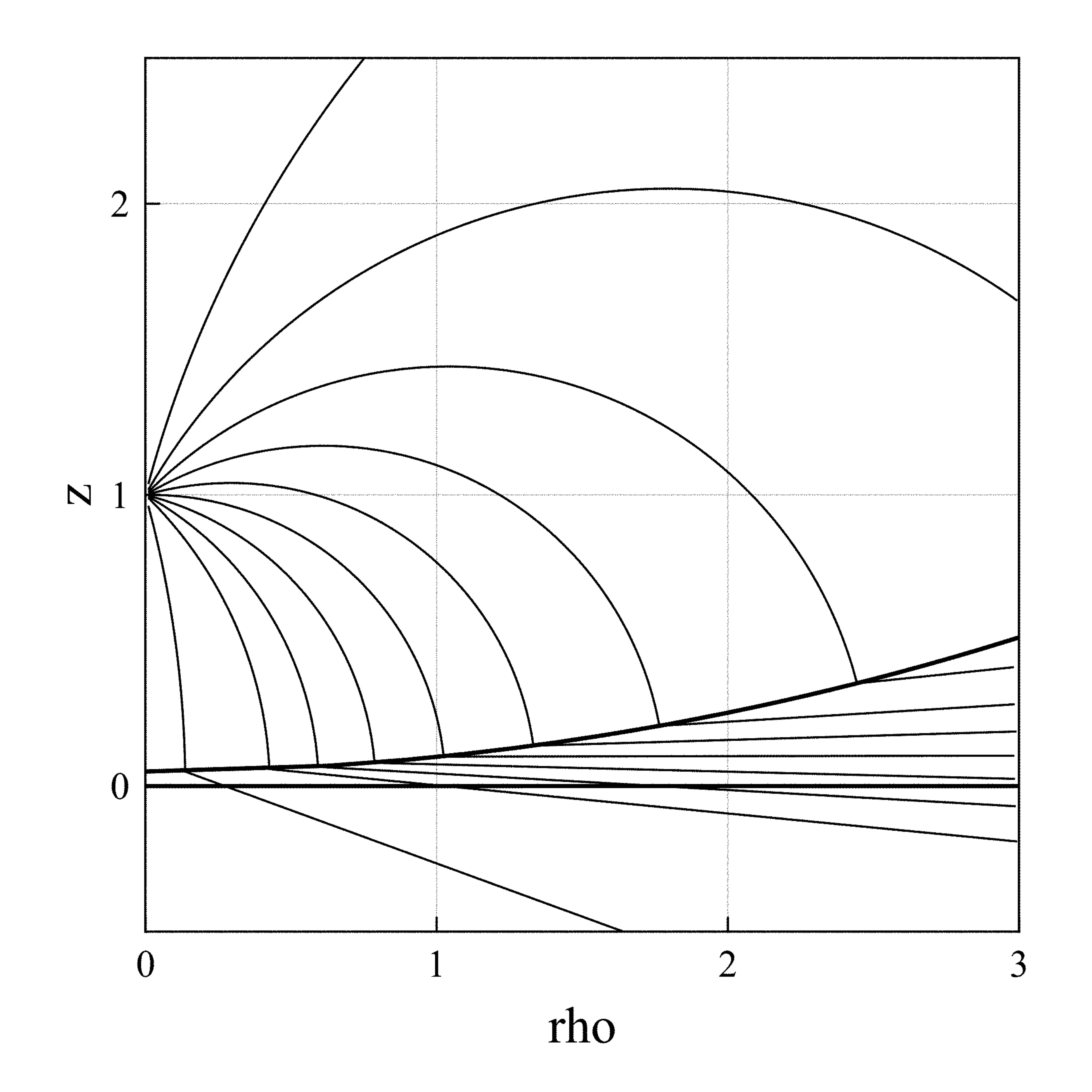}
\end{centering}
\caption{The same as Fig.~2, but with $T = -10$.}
\end{figure}

The break in the direction of the field lines in Fig.~3 is seen to be more extreme than that in Fig.~2, but the normal component of $\bf E$ is still continuous, and the discontinuity in $\bf E$ transverse is still compensated by the discontinuity in $\bf B$.
In contrast to our Fig.~3, the corresponding Fig.~4 in \cite{fg} is particularly unusual. That figure shows no relation between the constant velocity motion and the ensuing accelerated motion. A  ``connecting field" was introduced to somehow close the gap between these disconnected field lines.  But the connecting field  has no physical basis for this case.
\begin{figure}[h]
\begin{centering}
\includegraphics[height=4in]{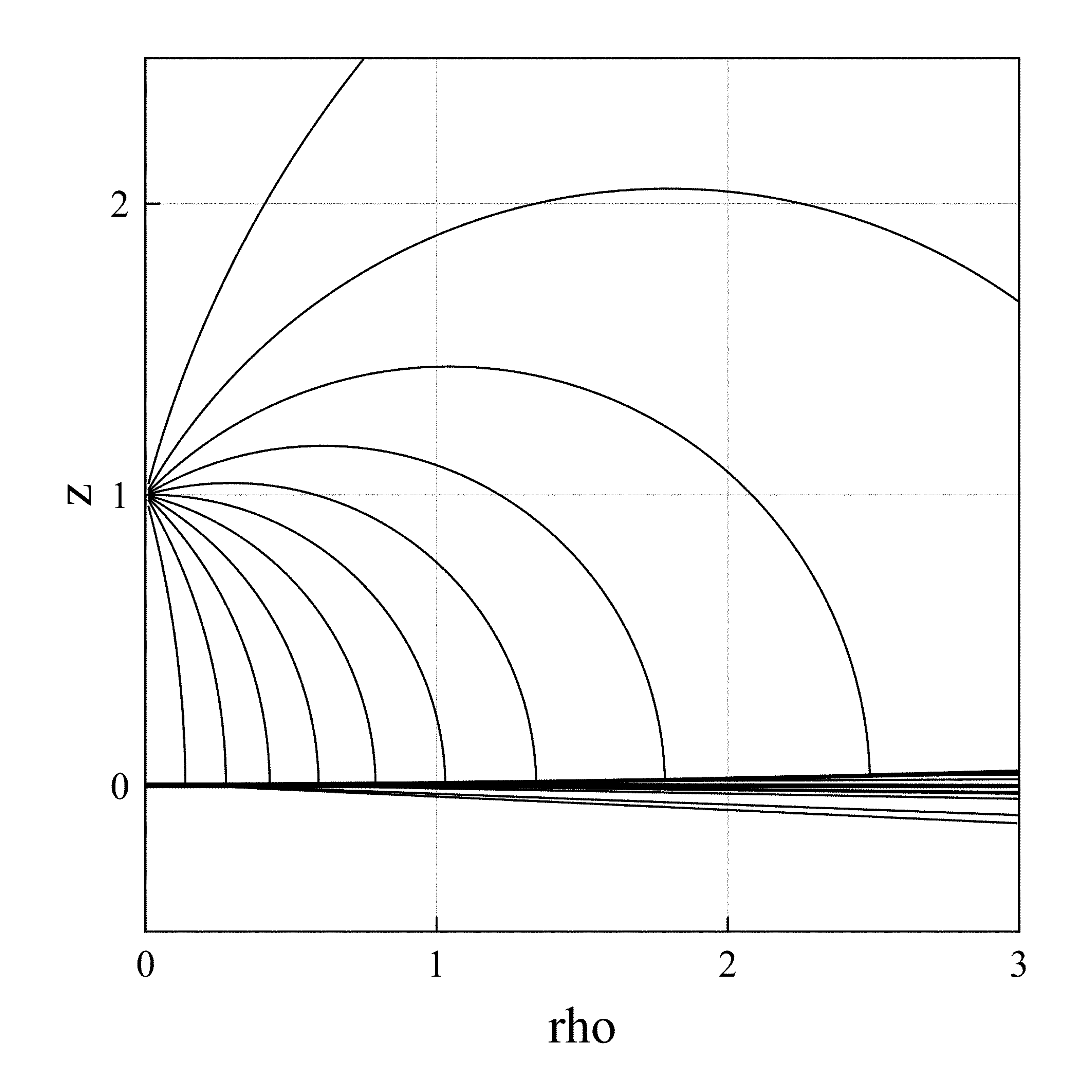}
\end{centering}
\caption{The same as Figs.~2 and 3, but with $T = -100$.}
\end{figure}
\newpage
Figure 4 has a very early transition time, $T = -100 z_0$, and shows field lines that look like Fig.~1 for positive $z $, but seem to be approaching a delta function near $z=0$.  The sequence of figures 2, 3, and 4 is seen to be progressing from an almost normal electric field to a combination of the field of Fig.~1 with the addition of a delta function along the z-axis.  

We can tell that the field lines are approaching a delta function along the z-axis for large negative transition times by looking at 
Eq.~(\ref{Er}) for the constant velocity part of the electric field as $Z\rightarrow \infty $.   
In this limit, the rho component, $E_\rho$, becomes infinite for  
$z=0$, and gets arbitrarily small for $z\ne 0$.  
At the same time, for $Z\rightarrow \infty $, 
the integral of $E_\rho$ from $z=-\infty$ to the spherical surface at $z=Z-\sqrt{T^2-\rho^2}$ 
is given by
\begin{eqnarray}
\lim_{z\rightarrow\infty}\int_{-\infty}^{Z-\sqrt{T^2-\rho^2}} E_\rho dz
&=&\lim_{z\rightarrow\infty}\int_{-\infty}^{(\rho^2+z_0^2)/2Z}\frac{\rho Zz_0^2 dz}{[\rho^2 z_0^2+(zZ-z_0^2)^2]^{3/2}}\nonumber\\
&=&\frac{2\rho}{\rho^2+z_0^2}.
\label{inte}
\end{eqnarray}
The integration in Eq.~(\ref{inte}) has to be over the entire range of the constant velocity electric field, from just at the transition sphere to $z=-\infty$, in order to get the correct factor for the delta function. 

We conclude from this result that the electric field emitted by a unit charge in truncated hyperbolic motion 
with the transition time $T\rightarrow -\infty$
is given by the acceleration field in Eq.~(\ref{Eh}) plus the delta function
 we have calculated above.  The result is
\begin{equation}
{\bf E}(\rho,z)=\frac{4z_0^2[(z^2-z_0^2-\rho^2){\bf{\hat z}}+2z\brho]}{[(\rho^2+z^2+z_0^2)^2
-(2zz_0)^2]^{\frac{3}{2}}}\Theta(z)
+\frac{2\brho}{\rho^2+z_0^2}\delta(z).
\label{Edelta}
\end{equation}
This is just the electric field that has been conjectured in \cite{fg},
 and derived by others\cite{bg,dgb,cross} using differing methods.\\

\noindent
{\large\bf ACKNOWLEDGEMENT}\\

I would like to thank David J. Griffiths and Daniel J. Cross for useful correspondence.

\end{document}